\documentstyle[aps,psfigure,epsfig]{revtex}

\begin{document}

\title{Helicity Signatures in Subthreshold $\rho^0$ Production on Nuclei}

\author{G.J.~Lolos$^1$\footnotemark, G.M.~Huber$^1$, A.~Shinozaki$^1$, Z.~Papandreou$^1$,  
E.J.~Brash$^1$, K.~Hossain$^1$, M.~Iurescu$^1$, D.~Nordin$^1$, G.~Garino$^2$,   
K.~Maruyama$^2$, K.~Maeda$^3$, T.~Suda$^3$, A.~Toyofuku$^3$, A.~Sasaki$^4$, 
H.~Yamashita$^5$}

\address{$^1$ Department of Physics, University of Regina, Regina, SK, S4S 0A2, 
Canada} 
\address{$^2$ Institute for Nuclear Study, University of Tokyo, Tanashi, Tokyo 
188, Japan}
\address{$^3$ Department of Physics, Tohoku University, Sendai 980, Japan}
\address{$^4$ College of General Education, Akita University, Akita 010, Japan}
\address{$^5$ Department of Applied Physics, Tokyo University of Agriculture 
and Technology, Koganei, Tokyo 184, Japan}

\author{(The TAGX collaboration)}
\date{\today}
\maketitle

\renewcommand{\thefootnote}{\fnsymbol{footnote}}
\footnotetext{* Contact author: E-mail: george.lolos@uregina.ca}

\begin{abstract}
We report a helicity analysis of subthreshold $\rho^0$ production on $^2H$
and $^{12}C$ nuclei at low photoproduction energies and large $|t|$. 
The results are indicative of a large longitudinal $\rho^0$ polarization
($l=1$, $m=0$) and are consistent with a strong helicity-flip mechanism of
$\rho^0$ production.  The analysis is model-independent and supports the large
mass shift reported from the earlier $^3He$ experiments.
\end{abstract}


At high densities and temperatures of matter, such as are present in
high-energy heavy-ion collisions, neutron stars and supernovae, particle
properties are expected to be modified from their free state (vacuum) values.
Because physical processes, such as those involved in the onset of chiral
symmetry restoration and the phase transition to quark gluon plasma, depend on
hadron properties in the nuclear medium, the role medium modifications play is
of significant importance \cite{Ko97}.

Heavy ion collider experiments have provided the first measurements of dilepton
invariant mass spectra which exhibit an enhancement at lower mass regions, over
that expected from vacuum mass values for the vector mesons \cite{CERES}.
Nevertheless, the complexity of the reaction has raised alternate and more
conventional explanations \cite{Dr96}.  Evidence for $\rho$ and $\omega$ mass
modification at normal nuclear matter densities has also been reported
\cite{Oz01}.

The first experiments utilizing tagged photons to probe medium effects have
been completed at INS-ES, using the large acceptance TAGX detector, and have
been reported in the literature \cite{Lo98,Hu98,Ka99}.  The results were
consistent with a substantial $\rho^0$ mass modification in the
$^3He(\gamma,\pi^+ \pi^-)ppn$ reaction.  The photon energy region was mostly
below the nominal 1080 MeV threshold on the free proton and was thus termed the
$subthreshold$ regime.  In this energy region, the Fermi momentum of the struck
nucleon was utilized to produce the $\rho^0$ via a quasi-free $\gamma-N$
interaction.  There were, however, mainly two reservations raised regarding
these conclusions: (1) the $\rho^0$ signal in the $\pi^+ \pi^-$ data-set is
small compared to competing processes (mainly $\Delta \pi$, $N^* \pi$, $N^*$,
and $\Delta \Delta$) and (2) the analysis is model- and assumption- dependent.
The spectral shape of the $\rho^0$ in the nuclear medium is not known and may
be quite different than the free spectral shape assumed in the fits.  Although
that analysis had been demonstrated to be relatively insensitive to reasonable
variations in the cross sections of the background processes and the width of
the $\rho^0$ \cite{Ka99}, a $\rho^0$ signature based on fewer assumptions
would increase the confidence in the interpretation of those results.  Such
analysis was first reported in \cite{Ka99} in the $J=1$ signature of the
$\rho^0$ in the $\rho^0 \rightarrow \pi^+ \pi^-$ decay.  In this letter, we
describe the helicity analysis for two nuclear targets and the implications on
the reaction mechanism and the mass modification.

The TAGX detector and the particle identification analysis used in the $^3He$
experiment have been described in detail elsewhere \cite{Ka99}.  Results from
$^{12}C$ and $CD_2$ targets -- in the form of research grade graphite and
deuterated polyethylene -- are reported for the first time here.  The
experimental techniques were the same as in \cite{Ka99} and will be presented
in more detail in a later publication.  Subtraction of the normalized $\pi^+
\pi^-$ spectra for these two targets isolated the contribution of the
deuterium.

At energies above 1.2 GeV, coherent (diffractive) production dominates the
$\rho^0$ photoproduction process ($|t| \le 0.1$ (GeV/c)$^2$).  At such low $t$,
the angular distribution of a decay pion with respect to the momentum vector of
the di-pion system, $\theta^*_{\pi^+}$, exhibits a sin$^2\theta^*_{\pi^+}$
dependence.  This is understood through helicity conservation: the $\rho^0$
carries the helicity of the transversely polarized photon.  Therefore, in the
$s$-channel helicity conserving frame, helicity conservation results in pure
sin$^2\theta^*_{\pi^+}$ distribution for unpolarized photons and the $\rho^0$
carries the photon's $l=1,$ $m=\pm 1$ quantum numbers.  Even though such
behaviour has been well established over a very wide photon energy range, there
have been small helicity violation effects, for example those reported by the
H1 collaboration \cite{H1}.  In contrast, the TAGX experiments are at much
lower energies and higher $t$, and display strong cos$^2\theta^*_{\pi^+}$
dependence in the $s$-channel helicity frame \cite{Ka99,Lo00}, which implies
that a helicity-flip mechanism dominates for those kinematics.

Among all $\pi^+\pi^-$ production processes in the 1 GeV photon energy region,
only $\rho^0$- and $\sigma$-meson decay can produce $\pi^+\pi^-$ with a
two-body kinematic distribution and a $\pi\pi$ opening angle of $180^o$ on an
event-by-event basis in their respective rest frames.  All other known
processes are two-step in nature, with the two pions generated at different
production vertices.  The two-pion kinematical observables will follow
multi-body phase-space and the kinematic correlation between the two pions will
be weaker.  This is a simple argument based solely on kinematic considerations
and is independent of any relative angular momentum considerations among the
two-step reaction products.

Returning to the single-step $\rho\rightarrow\pi\pi$ process, in the
subthreshold region the Lorentz boost of the $\rho^0$ in the lab frame is
small and the $\pi^+\pi^-$ opening angle remains large.  All other reactions
leading to $\pi^+\pi^-$ production result in significantly different angular
correlations.  This has been demonstrated with the use of Monte Carlo (MC)
simulations incorporating the TAGX geometry and trigger requirements and has
been reported in \cite{Ka99,Lo00}.  Given the strong correlation, only a small
fraction of the pions originating from the $\rho^0 \rightarrow \pi^+ \pi^-$
decay is eliminated by an opening angle requirement of $\theta_{\pi \pi} \ge
120^o$.  However, a large fraction of $\pi^+ \pi^-$ events originating from
reactions other than $\rho^0$ and $\sigma$ decays is eliminated.

\begin{figure}
    \centerline{\epsfig{file=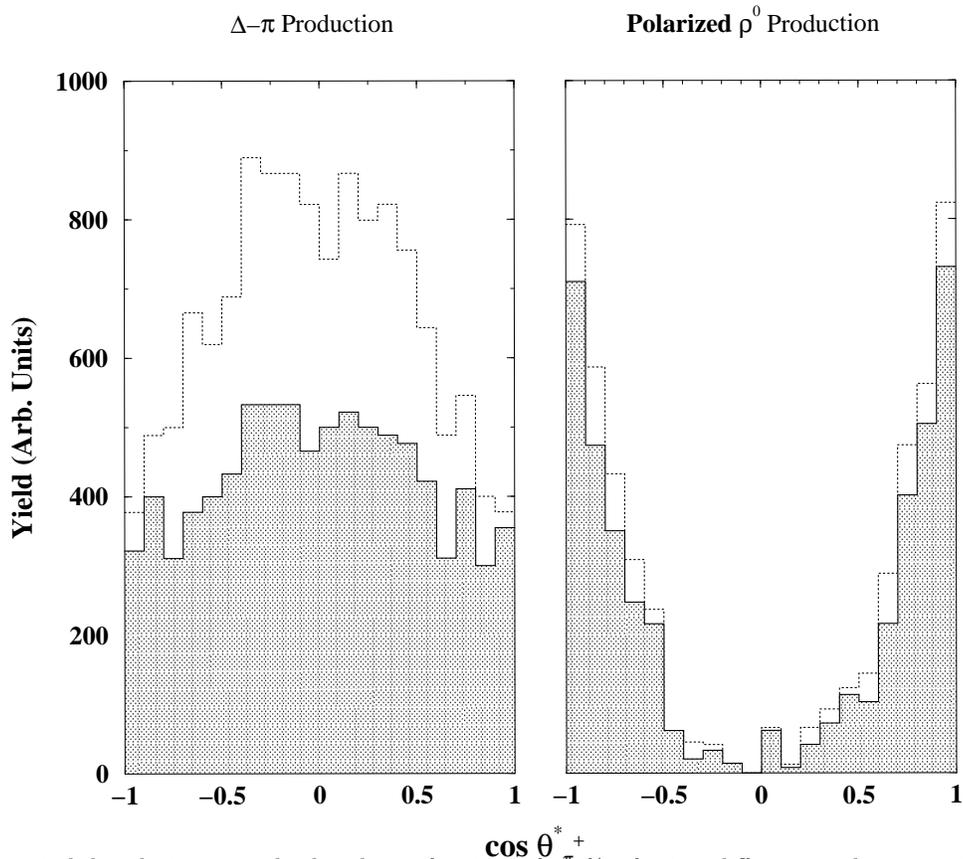,height=5.in,angle=-89.9}}
\caption{MC generated distributions are displayed as a function of
cos$\theta^*_{\pi^+}$ for two different mechanisms at $E_{\gamma}=700-1000$
MeV.  The cuts applied are $\theta_{\pi \pi}\ge 0^o$ (upper curve) and
$\theta_{\pi \pi} \ge 120^o$ (shaded area), respectively.}
\end{figure}

Fig. 1 demonstrates the effect of the $\theta_{\pi\pi}>120^0$ lab frame opening
angle cut upon two of the several mechanisms contributing to the data-set.  The
modeled processes shown are based on the quasi-free production reactions
$\gamma p_{F} \rightarrow \Delta^{++} \pi^- n$, and $\gamma p_{F} \rightarrow
\rho^0 p'$, where $p_{F}$ denotes a bound proton with the single nucleon Fermi
momentum distribution appropriate for the $^2H$ and $^{12}C$ nuclei.  The
processes shown in Fig.1 are representative of the MC studies carried out in
Refs. \cite{Ka99,Lo00} for the dominant channels contributing to $\pi^+ \pi^-$.
The cos$^2\theta^*_{\pi^+}$ distribution exhibited in the right panel can only
be produced by longitudinally polarized $\rho^0$ ($l=1,m=0$) decay; all other
processes give distributions similar to that shown in the left panel.
Therefore, a definitive signature of $\rho^0$ content in the data is found at
kinematical regions which are little affected by such an opening angle cut.

The helicity angle $\theta^*_{\pi^+}$ is defined in the di-pion rest system in
the following manner: (i) the summed momentum vector $\vec{p}_{\pi \pi}$ is
formed, (ii) the longitudinal component $p_{\pi^+//}$ of the $\pi^+$ momentum
with respect to $\vec{p}_{\pi \pi}$ in the lab frame is calculated; (iii) both
$p_{\pi^+}$ and $p_{\pi^+//}$ are transformed to the $\pi^+\pi^-$ center of
mass frame, and $\theta^*_{\pi^+}$ is calculated from $\rm
cos\theta^*_{\pi^+}=p_{\pi^+//} / p_{\pi^+}$.

If one accepts the existence of the $\sigma$-meson, the two pions produced as a
result of its decay will also have opening angles of $180^o$ on an
event-by-event basis in the $\sigma$ rest frame.  However, since the width of
the $\sigma$ is much broader than that of the $\rho^0$ \cite{pdg}, $\sigma$
decay will effectively populate all the phase space acceptance available to the
experiment and the Lorentz boost values will thus occupy a broader region than
$\rho^0$ decay.  This will result in a broader distribution of the pion opening
angles.  Therefore, when one considers only opening angle and
$\theta^*_{\pi^+}$ distributions, the experimental signature of the $\sigma$
will be nearly indistiguishable from that of the $unpolarized$ $\rho^0$.  The
$\sigma$ cannot, however, differentially populate the regions of interest with
$\rm |cos\theta^*_{\pi^+} \ge 0.5|$, which are characteristic of the polarized
$\rho^0$, due to the $\sigma$-meson's scalar ($J=0$) nature.  In conclusion, no
other $\pi^+ \pi^-$ production reaction has been identified which will have
$both$ the $p-wave$ ($l=1,m=0$) signature of the polarized $\rho^0$ decay $and$
the two-body decay opening angles of the $\rho^0$. The $\rho^0 \rightarrow
\pi^+ \pi^-$ decay uniquely satisfies both conditions.

The $\pi^+\pi^-$ data are displayed in Figures 2 and 3 for the $^2He$ and
$^{12}C$ targets, respectively, in the $E_{\gamma}=700-1000$ MeV tagged photon
energy region.  In Fig. 2, the flatness of the distribution for the data-points
with no $\theta_{\pi\pi}$ cut is characteristic of the superposition of
background processes dominating the region between $\sim \pm 0.5$.  The shaded
distribution, subject to $\theta_{\pi\pi}>120^o$, clearly shows that the
populations outside that region, and near the edges of the distribution, are
affected very little by the application of the $\theta_{\pi \pi}$ cut.
However, in the region within $\pm 0.5$, the population is reduced by a factor
of five.  The $p-wave$ shape ($l=1,m=0$) is clearly observed.  The drop in the
population at cos$\theta^*_{\pi^+}=\pm 1$ in Fig. 2 is due to the TAGX
acceptance and has been verified by MC simulations.  The application of a
missing mass cut in panel (b) further suppress multi-pion production.  The loss
of events is verified by a visual comparison of panels (a) and (b).  The
$p-wave$ signature in the shaded distribution is further enhanced by the
missing mass cut, since multi-pion production cannot have the kinematic
signatures of a two body decay.

\begin{figure}
    \centerline{\epsfig{file=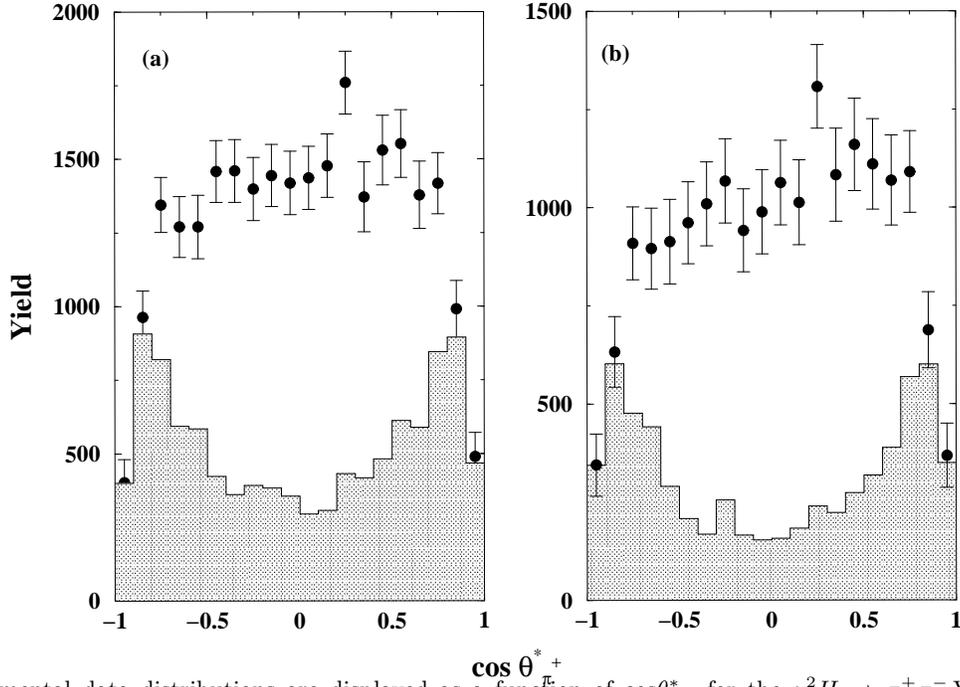,height=5.in,angle=-89.9}}
\caption{Experimental data distributions are displayed as a function of
cos$\theta^*_{\pi^+}$ for the $\gamma ^2H \rightarrow \pi^+ \pi^- X$ reaction
and for $E_{\gamma}=700-1000$ MeV.  Panel (a): $\theta_{\pi \pi}\ge 0^o$
(points) and $\theta_{\pi \pi} \ge 120^o$ (shaded area) cuts applied,
respectively.  Panel (b): the corresponding distributions from panel (a)
but with an additional missing mass cut applied, $1725 \le m_{miss} \le 2000$
MeV/c$^2$.}
\end{figure}

Similar observations hold for the $^{12}C$ target as well, as shown in Fig. 3.
The statistical quality of the carbon data is not as good as that for the $^2H$
data due to the thinness of the graphite target, however, the effect of the
opening angle cut in both panels (a) and (b) is as dramatic as in
Fig. 2.  There is also no doubt that the distributions share very similar
features for both nuclei.  One also notes the less pronounced loss of
acceptance at the extreme values of the distribution, unlike the case of $^2H$
in Fig. 2.  This is due to the larger single nucleon Fermi momentum in carbon
and the resulting larger phase-space available to the $\pi \pi$ system.

The establishment of a unique signature for the $\rho^0$ in the TAGX $\pi^+
\pi^-$ data-set becomes a powerful tool with which to re-investigate the mass
modifications reported already in the literature \cite{Lo98,Hu98,Ka99} for
$^3He$.  However, one has to first consider the issues arising from this
helicity signature ($l=1,m=0$) for the $\rho^0$.  Small helicity violations,
such as reported in \cite{H1} (and elsewhere), have been observed in high
virtuality (high $Q^2$) $\rho^0$ lepto-production data.  Strong and rapidly
rising helicity-flip amplitudes had also been observed in the $t$-channel,
using photons, for values of $|t| \ge 0.4$ (GeV/c)$^2$ \cite{Ba72}.  In this
work, the $|t|$ values range between 0.6 and 1.1 (GeV/c)$^2$ and exceed the
values in \cite{Ba72}.  Unfortunately, the limited statistics available to the
analysis do not allow a finer binning in $t$ and in invariant mass to further
investigate the correlation.  It has been verified, within the limited
statistics, that the $\rho^0$ candidates from regions of high $t$ are
correlated to $\rho^0$ candidates in the $|cos\theta^*_{\pi^+}|\ge 0.7$
distributions.  One explanation proposed for such strong helicity-flip is based
on a topological feature of QCD and is more applicable in the low energy regime
\cite{Bo00}.  It has its foundation in the coupling of the $\pi^0$ exhange
particle to the $\gamma$ and $\pi^+ \pi^-$ vertex.  It remains to be seen if
such a mechanism accounts for the features and cross sections which will result
from this work.

\begin{figure}
    \centerline{\epsfig{file=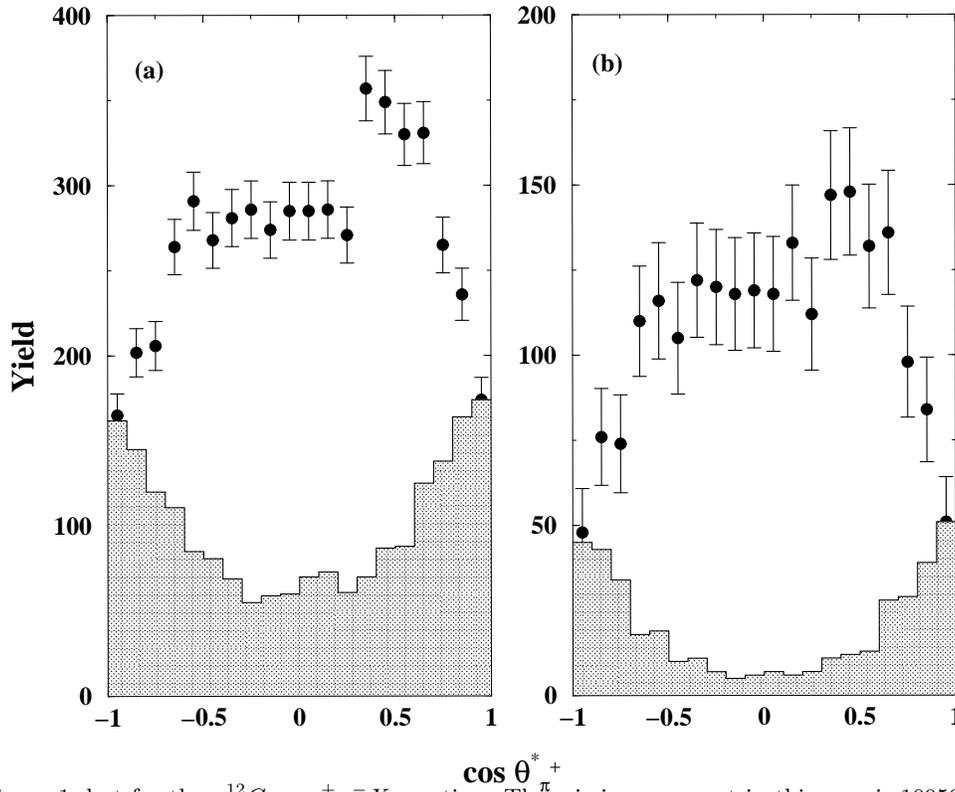,height=5.in,angle=-89.9}}
\caption{As in figure 1, but for the $\gamma ^{12}C \rightarrow \pi^+ \pi^- X$
 reaction.  The missing mass cut in this case is $10950 \le m_{miss} \le 11300$
 MeV/c$^2$.}
\end{figure}

Data distributions for $\pi^+\pi^-$ production from $^2H$, in the $700\le
E_{\gamma}\le 1000$ MeV range, are shown in Fig. 4 for four $\pi\pi$ invariant
mass ($m_{\pi \pi}$) regions.  One observes a pronounced $p-wave$-like
$(l=1,m=0)$ signature in the $600 \le m_{\pi \pi} \le 700$ MeV/c$^2$ region.
The 500-600 MeV/c$^2$ $m_{\pi \pi}$ bin exhibits a weaker $p-wave$ signature
and the result of the opening angle cut is a dramatic reduction ($\sim 10:1$)
in the number of surviving events in the central region.  This is consistent
with the MC simulations that point to the dominant role the $\Delta \pi$ and
$\Delta \Delta$ production processes play in this invariant mass range
\cite{Ka99,Lo00}.  On the other hand, in this $E_{\gamma}$ energy region,
little strength for the $\rho^0$ is observed in the ``natural'' $m_{\pi \pi}$
range of 700-800 MeV/c$^2$.  Even in this case, however, $\rho^0$ production is
associated with a helicity-flip signature, from the $l=1,m=\pm 1$ substate
dominant in coherent production to the $l=1,m=0$ substate mostly associated
with the higher $t$ production mechanism in this experiment.

By assuming a slowly varying background in the invariant mass regions of
400-800 MeV/c$^2$, an assumption borne out by similar analyses in
\cite{Ka99,Lo00}, the shaded areas are consistent with an effective mass of the
$\rho^0$ populating the 500-700 MeV/c$^2$ region.  Phase-space acceptance
corrections accounting for the overlap between the actual $\rho^0$ spectral
function, the tagged photon beam energy spectrum and the photon energy
distributions may result in a modest upward correction, but, nevertheless, this
assumption-free analysis is consistent with a significantly lower $\rho^0$ {\it
effective mass}.  The results are not strongly supporting of a much broader
$\rho^0$ spectral shape or one with a long, low-invariant-mass tail of any
significance, at this time.

\begin{figure}
    \centerline{\epsfig{file=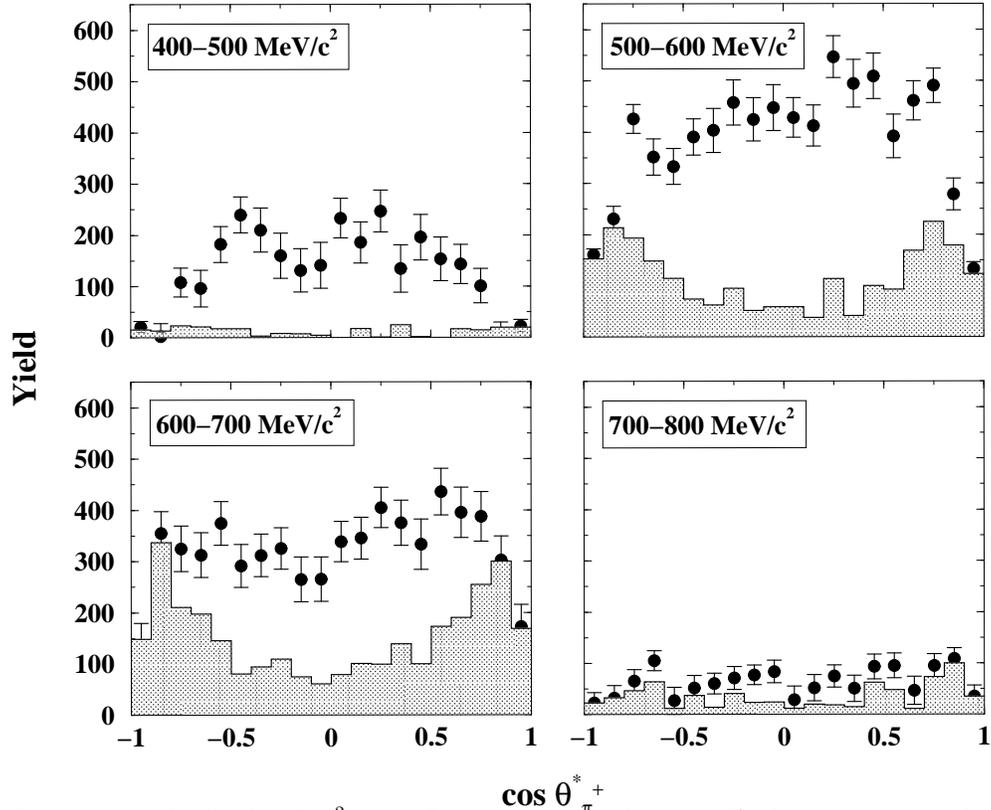,height=5.in,angle=-89.9}}
\caption{Experimental data distributions for $^2H$ are displayed as a function
of cos$\theta^*_{\pi^+}$ in the four panels in the invariant mass range of
400-800 MeV/c$^2$ accessible to the experiment and in 100 MeV/c$^2$ bins.  All
data in this figure have missing mass cuts applied, as explained in the text.
The data points with the error bars are the total data without any $\theta_{\pi
\pi}$ cut applied, while the shaded areas are the result of a $\theta_{\pi \pi}
\ge 120^o$ cut.}
\end{figure}

A comparison of the data in the 700-800 MeV/c$^2$ bin to the distribution and
population in the 600-700 MeV/c$^2$ bin is consistent with a much weaker
$\rho^0$ signature in the former as compared to the latter, of near
vacuum-value $m_{\rho^0}$ but also with a helicity-flip signature.  In this
higher $m_{\pi \pi}$ bin, however, some loss of strength can be accounted for
by phase-space corrections and photon yield normalizations.  However, even such
corrections cannot account for the much higher population evident in the
600-700 MeV/c$^2$ mass bin, unless the main strength of the $\rho^0$ mass
distribution is confined in the latter invariant mass range.  It is worth
mentioning here that a similar conclusion, for the break-up channel in $^3He$,
has been reached by a completely different analysis and reported in
\cite{Ma00}.  A more detailed analysis taking into account various modified
spectral shapes for the $\rho^0$, corrections for phase-space acceptance, as
well as photon yields and spectral distributions, is underway to further
investigate this behaviour and will be reported at a later date.

In conclusion, $\pi^+ \pi^-$ photoproduction on light nuclei in the
$E_{\gamma}$ energy region of 700 to 1000 MeV, which lies mostly below the
$\rho^0$ production threshold on the free proton, is reported here with
emphasis on helicity information.  The large values of $t$ in this work confirm
similar helicity-flip evidence reported earlier and at higher energies
\cite{Ba72}.  The helicity analysis, based only on the kinematical signatures
expected in the decay of a particle into two pions, is consistent with a
particle with an invariant mass distribution contained mostly within a 500-700
MeV/c$^2$ region and with a dominant helicity zero ($l=1,m=0$) state.  In the
absence of any known particle with such quantum numbers and mass, the data are
consistent with the production of the $\rho^0$ and, thus, they imply strong
helicity-flip amplitudes in the reaction mechanism.  Such a unique signature of
$\rho^0$ in the data-set potentially allows the reliable separation of $\rho^0$
from unrelated background.  The strength and symmetry of the $l=1,m=0$
signatures in the data support our earlier reports of a decrease of the {\it
effective mass} of the $\rho^0$.  Furthemore, these conclusions are largely
independent of nuclear target and appear to emphasize a common underlying
mechanism of the $\gamma -N$ interaction, in the nuclear environmnt, leading to
$\rho^0$ production. 

The authors wish to thank the staff of INS-ES for their hospitality and help    
during the experiments.  This work has been partially supported by grants in
aid of research by NSERC and INS-ES.


\begin{references}

\bibitem{Ko97} C.M. Ko, V. Koch and G. Li, Annu. Rev. Nucl. Part. Sci. {\bf 47},
505 (1997)
\bibitem{CERES} CERES Collaboration, Th. Ulrich {\it et al.}, Nucl. Phys. {\bf 
A610}, 310 (1996).
\bibitem{Dr96} A. Drees, Nucl. Phys. {\bf A610}, 535 (1996).
\bibitem{Oz01} K. Ozawa {\it et al.}, nucl-ex/0011013.
\bibitem{Lo98} G.J. Lolos {\it et al.}, Phys. Rev. Lett. {\bf 80}, 241 (1998).
\bibitem{Hu98} G.M. Huber {\it et al.}, Phys. Rev. Lett. {\bf 80}, 5285 (1998).
\bibitem{Ka99} M.A. Kagarlis {\it et al.}, Phys. Rev. C {\bf 60}, 025203 
(1999).
\bibitem{H1} B. Clerbaux, hep-ph/9905507;hep-ph/9908519.
\bibitem{Lo00} G.J. Lolos, Hadrons in Dense Matter. Proceedings of the International Workshop XXVIII on
Gross Properties of Nuclei and Nuclear Excitations, Hirschegg, Austria, 2000. 
Edited by M. Buballa {\it et al.}, nucl-ex/0101005.
\bibitem{pdg} Particle Data Group, Eur. Phys. J. C {\bf 15}, 1 (2000).
\bibitem{Ba72} J. Ballam {\it et al.}, Phys. Rev. D {\bf 5}, 545 (1972).
\bibitem{Bo00} D. Boer, D. Kharzeev, and R.D. Pisaski, private commuication
and in preparation.
\bibitem{Ma00} K. Maruyama, Proceedings of the Second KEK-Tanashi International
Symposium on Hadron and Nuclear Physics with Electromagnetic Probes, Tokyo,
October 1999. Edited by K. Maruyama and H. Okuno, published by Elsevier (2000).
\end{references}
\end{document}